\documentclass[12pt]{article}
\usepackage{amsmath}
\usepackage{amssymb}
\usepackage{amsfonts}

\numberwithin{equation}{section}
\newcommand{\field}[1]{\mathbb{#1}}
\newcommand{\R}{\field{R}}
\newcommand{\ap}{a^{\dagger}}
\newcommand{\bp}{b^{\dagger}}
\newcommand{\bb}{\bar{b}}
\newcommand{\bbp}{\bar{b}^{\dagger}}

\title{
${\cal N}=2$ supersymmetric extension of the Tremblay-Turbiner-Winternitz Hamiltonians on a plane}
\author{C Quesne\\ 
{\small Physique Nucl\'eaire Th\'eorique et Physique Math\'ematique,  Universit\'e Libre de Bruxelles,} \\ 
{\small Campus de la Plaine CP229, Boulevard~du Triomphe, B-1050 Brussels, Belgium}}
\date{ }
\begin{document}
\baselineskip=22pt plus 1pt minus 1pt
%%%%%%%%%%%%%%%%%%%%%%%%%%%%%%%%%%%%%%%%%%%%%%%%%%%%%%%%%%
\maketitle

\begin{abstract} 
The family of Tremblay-Turbiner-Winternitz Hamiltonians $H_k$ on a plane, corresponding to any positive real value of $k$, is shown to admit a ${\cal N} = 2$ supersymmetric extension of the same kind as that introduced by Freedman and Mende for the Calogero problem and based on an ${\rm osp}(2/2, \R) \sim {\rm su}(1,1/1)$ superalgebra. The irreducible representations of the latter are characterized by the quantum number specifying the eigenvalues of the first integral of motion $X_k$ of $H_k$. Bases for them are explicitly constructed. The ground state of each supersymmetrized Hamiltonian is shown to belong to an atypical lowest-weight state irreducible representation.
\end{abstract}


%
%========================================================================
%
\newpage
\section{Introduction}

Recently an infinite family of exactly solvable quantum Hamiltonians on a plane
\begin{equation}
\begin{split}
  & H_k = - \partial_r^2 - \frac{1}{r} \partial_r - \frac{1}{r^2} \partial_{\varphi}^2 + \omega^2 r^2 +
       \frac{k^2}{r^2} [a(a-1) \sec^2 k\varphi + b(b-1) \csc^2 k\varphi], \\
  & 0 \le r < \infty, \qquad 0 \le \varphi < \frac{\pi}{2k},
\end{split}  \label{eq:TTW}
\end{equation}
introduced by Tremblay, Turbiner and Winternitz (TTW) \cite{tremblay09}, has aroused a lot of interest because of the conjectured superintegrability of the Hamiltonians for all positive integer values of $k$. For the corresponding classical systems, it has been shown that all bounded trajectories are closed and that the motion is periodic for all integer and rational values of $k$ \cite{tremblay10}. Such classical systems have been proved to be superintegrable \cite{kalnins09} and generalizable to higher dimensions \cite{kalnins10a}. For the quantum Hamiltonians, the validity of the superintegrability conjecture has been demonstrated for any odd integer $k$ \cite{cq10a} by using a $D_{2k}$ extension of $H_k$ \cite{cq10b} and a Dunkl operator formalism previously employed in the $k=3$ case \cite{cq95}. Furthermore, a canonical operator method has been applied to provide a constructive proof that all quantum Hamiltonians with rational $k$ are superintegrable \cite{kalnins10b}.\par
%
%----------------------------------------------------------------------------------------------------------------------
%
Another attractive propery of the TTW family (\ref{eq:TTW}) is that it includes as special cases several well-known Hamiltonians, which have been much studied in the literature both for their intrinsic mathematical properties and for their possible physical applications. They correspond to the Smorodinsky-Winternitz (SW) system ($k=1$) \cite{fris, winternitz}, the rational $BC_2$ model ($k=2$) \cite{olsha81, olsha83} and the three-particle Calogero model \cite{calogero69} with some extra three-body interaction ($k=3$), initially introduced by Wolfes and also considered by Calogero and Marchioro (CMW model) \cite{wolfes, calogero74}. The fact that the $k=2$ and $k=3$ cases belong to the family of Calogero-type Hamiltonians associated with root systems of classical Lie algebras \cite{olsha81, olsha83} has surely contributed to the interest aroused by the TTW Hamiltonians.\par
%
%--------------------------------------------------------------------------------------------------------------
%
Some years ago, a ${\cal N}=2$ supersymmetric extension of the $\nu$-particle Calogero model, based on a dynamical ${\rm osp}(2/2, \R) \sim {\rm su}(1,1/1)$ superalgebra \cite{nahm, scheunert77a, scheunert77b, balantekin, frappat}, has been introduced by Freedman and Mende \cite{freedman}. Several aspects of this extension and of its generalization to other root systems than that of the $A_{\nu-1}$ Lie algebra have been reviewed in the literature (see, e.g., \cite{brink93, brink98, ghosh01}). These supermodels are exactly solvable and play an important role in many areas of physics, such as the study of superstrings, black holes, superconformal quantum mechanics, spin chains, etc.\par
%
%-----------------------------------------------------------------------------------------------------------------
%
Since then, there have been great theoretical advances in that field too. For instance, the nonuniqueness of the construction proposed by Freedman and Mende has been discussed \cite{ghosh04, gala06}. Calogero-like models have also been analyzed in terms of hidden nonlinear supersymmetries \cite{plyu96, plyu00, correa}. Furthermore, some ${\cal N}=4$ supersymmetric extensions have been recently considered in connection with the ${\rm su}(1,1/2)$ superalgebra \cite{wyllard, gala07, krivonos, gala09} or more generally with $D(2,1;\alpha)$ \cite{hakobyan}.\par
%
%-----------------------------------------------------------------------------------------------------------
% 
The purpose of the present paper is to show that the family of TTW Hamiltonians corresponding to any positive real value of $k$ admits a ${\cal N}=2$ supersymmetric extension including the standard one of Calogero-type Hamiltonians \cite{freedman, brink93, brink98, ghosh01} as special case for $k=2$ and 3.\par
%
%------------------------------------------------------------------------------------------------------------------
%
In section 2, we review some known realizations of the ${\rm osp}(2/2, \R)$ superalgebra in cartesian coordinates. In section 3, we make a transformation to polar coordinates on a plane to deal with the case of the TTW Hamiltonians. Bases for the corresponding irreducible representations (irreps) are constructed in section 4. The $k=1$, 2 and 3 examples are treated in detail in section 5. Finally, section 6 contains the conclusion.\par
%
%===================================================================
%
\section{\boldmath Realizations of the ${\rm osp}(2/2, \R)$ superalgebra}

The ${\rm osp}(2/2, \R)$ superalgebra is generated by eight operators, four even ones closing the ${\rm sp}(2, \R) \times {\rm so}(2)$ Lie algebra and four odd ones, which separate into two ${\rm sp}(2, \R)$ spinors \cite{nahm, scheunert77a, scheunert77b, balantekin, frappat}. In the Cartan-Weyl basis, the former can be written as $K_0$, $K_{\pm}$ and $Y$, while the latter are denoted by $V_{\pm}$ and $W_{\pm}$. They satisfy the following (nonvanishing) commutation or anticommutation relations
\begin{equation}
\begin{array}{ll}
  [K_0, K_{\pm}] = \pm K_{\pm}, &\qquad  [K_+, K_-] = - 2K_0, \\[0.3cm]
  [K_0, V_{\pm}] = \pm \tfrac{1}{2} V_{\pm}, &\qquad [K_0, W_{\pm}] = \pm \tfrac{1}{2} W_{\pm}, 
  \\[0.3cm]
  [K_{\pm}, V_{\mp}] = \mp V_{\pm}, &\qquad [K_{\pm}, W_{\mp}] = \mp W_{\pm}, \\[0.3cm]
  [Y, V_{\pm}] = \tfrac{1}{2} V_{\pm}, &\qquad [Y, W_{\pm}] = - \tfrac{1}{2} W_{\pm}, \\[0.3cm] 
  \{V_{\pm}, W_{\pm}\} = K_{\pm}, &\qquad \{V_{\pm}, W_{\mp}\} = K_0 \mp Y,      
\end{array}  \label{eq:def-1}
\end{equation}
together with the Hermiticity properties
\begin{equation}
  K_0^{\dagger} = K_0, \qquad K_{\pm}^{\dagger} = K_{\mp}, \qquad Y^{\dagger} = Y, \qquad 
  V_{\pm}^{\dagger} = W_{\mp}.  \label{eq:def-2}
\end{equation}
From (\ref{eq:def-1}), it is clear that $K_0$ and $Y$ are the weight generators, while $K_-$, $V_-$, $W_-$ (resp.\ $K_+$, $V_+$, $W_+$) are the lowering (resp.\ raising) generators.\par
%
%-----------------------------------------------------------------------------------------------------------------
%
A well-known realization of this superalgebra uses $\nu$ commuting pairs of bosonic and fermionic creation and annihilation operators, $\ap_i$, $a_i$ and $\bp_i$, $b_i$, $i=1$, 2,~\ldots, $\nu$, where $[a_i, \ap_j] = \delta_{ij}$ and $\{b_i, \bp_j\} = \delta_{ij}$, respectively. Such a realization is given by
\begin{equation}
\begin{split}
  & K_0 = \frac{1}{2} \biggl(\sum_i \ap_i a_i + \frac{\nu}{2}\biggr), \qquad K_+ = \frac{1}{2} \sum_i 
        a^{\dagger2}_i, \qquad K_- = \frac{1}{2} \sum_i a_i^2,  \\
  & Y = \frac{1}{2} \biggl(\sum_i \bp_i b_i - \frac{\nu}{2}\biggr), \qquad V_+ = \frac{1}{\sqrt{2}} 
        \sum_i \ap_i \bp_i, \qquad V_- = \frac{1}{\sqrt{2}} \sum_i a_i \bp_i,  \\ 
  & W_+ = \frac{1}{\sqrt{2}} \sum_i \ap_i b_i, \qquad W_- = \frac{1}{\sqrt{2}} \sum_i a_i b_i, 
\end{split}  \label{eq:oscillator}
\end{equation}
where all summations run over 1, 2,~\ldots, $\nu$. It is related to the superoscillator (see, e.g., \cite{bagchi}), for which
\begin{equation}
  {\cal H}^s = 4 \omega (K_0 + Y), \qquad Q = 2 \sqrt{\omega}\, W_+, \qquad Q^{\dagger} = 2 \sqrt{\omega}\,
  V_-  \label{eq:SUSY}
\end{equation}
provide a realization of the ${\rm sl}(1/1)$ superalgebra of standard supersymmetric quantum mechanics,
\begin{equation*}
  [{\cal H}^s, Q] = [{\cal H}^s, Q^{\dagger}] = 0, \qquad \{Q, Q^{\dagger}\} = {\cal H}^s,
\end{equation*}
and the operators (\ref{eq:oscillator}) generate a dynamical superalgebra.\par
%
%----------------------------------------------------------------------------------------------------------
%
The boson-fermion realization (\ref{eq:oscillator}) can be generalized by including an additional contribution in the bosonic operators $\ap_i = (- \partial_i + \omega x_i)/\sqrt{2\omega}$, $a_i = (\partial_i + \omega x_i)/\sqrt{2\omega}$ (with $\partial_i \equiv \partial/\partial x_i$) appearing in the odd generators $V_{\pm}$, $W_{\pm}$. The latter become
\begin{equation}
  V_{\pm} = \frac{1}{2\sqrt{\omega}} \sum_i (\mp \partial_i + \omega x_i \mp \partial_i W) \bp_i, \qquad
  W_{\pm} = \frac{1}{2\sqrt{\omega}} \sum_i (\mp \partial_i + \omega x_i \pm \partial_i W) b_i,
  \label{eq:Calogero-1}
\end{equation}
where $W$ denotes some function of the $x_i$'s. Then the even operators
\begin{equation}
\begin{split}
  & K_0 = K_{0,{\rm B}} + \Gamma, \qquad K_{\pm} = K_{\pm,{\rm B}} - \Gamma, \qquad Y = \frac{1}{4}
       \sum_i (2x_i \partial_i W + [\bp_i, b_i]), \\
  & K_{0,{\rm B}} = D + \frac{1}{4} \omega \sum_i x_i^2, \qquad K_{\pm,{\rm B}} = - D
       + \frac{1}{4} \omega \sum_i x_i^2 \mp \frac{1}{4} \sum_i (2x_i \partial_i + 1), \\
  & D = \frac{1}{4\omega} \sum_i [- \partial_i^2 - \partial_i^2 W + (\partial_i W)^2], \qquad \Gamma =
       \frac{1}{2\omega} \sum_{ij} \partial_{ij}^2 W\, \bp_i b_j,
\end{split}  \label{eq:Calogero-2}
\end{equation}
resulting from the anticommutation relations in (\ref{eq:def-1}), also satisfy the remaining defining relations of ${\rm osp}(2/2, \R)$ provided
\begin{equation}
  \biggl[\sum_i x_i \partial_i, D\biggr] = - 2D, \qquad \biggl[\sum_i x_i \partial_i, \Gamma\biggr] = - 2\Gamma, 
  \qquad \sum_i x_i \partial_i W = C,  \label{eq:conditions}
\end{equation}
where $C$ is some constant. This is the kind of realization that leads to the dynamical superalgebra of Calogero-type Hamiltonians \cite{freedman, brink93, brink98, ghosh01} if $x_i$, $i=1$, 2,~\ldots, $\nu$, denote the coordinates of $\nu$ particles on a line and $W$ is an appropriate solution of equation (\ref{eq:conditions}).\par
%
%====================================================================
%
\section{\boldmath Realization of ${\rm osp}(2/2, \R)$ in polar coordinates for the TTW Hamiltonians}

Let us consider the case where there are only two variables $x_1$, $x_2$, which are the cartesian coordinates $x$, $y$ of a particle on a plane, and consequently two pairs of fermionic creation and annihilation operators, denoted by $(\bp_x, b_x)$ and $(\bp_y, b_y)$. On setting $x = r \cos \varphi$, $y = r \sin \varphi$, $W$ becomes a function of $r$, $\varphi$, and equations (\ref{eq:Calogero-1}), (\ref{eq:Calogero-2}) and (\ref{eq:conditions}) are transformed into
\begin{equation}
\begin{split}
  V_{\pm} & = \frac{1}{2\sqrt{\omega}} \biggl[\left(\mp \cos \varphi \partial_r  \pm \frac{1}{r} \sin \varphi
        \partial_{\varphi} + \omega r \cos \varphi \mp \cos \varphi \partial_r W \pm \frac{1}{r} \sin \varphi
        \partial_{\varphi} W\right) \bp_x \\
  & \quad + \left(\mp \sin \varphi \partial_r  \mp \frac{1}{r} \cos \varphi \partial_{\varphi} + \omega r 
        \sin \varphi \mp \sin \varphi \partial_r W \mp \frac{1}{r} \cos \varphi \partial_{\varphi} W\right) \bp_y 
        \biggr], \\
  W_{\pm} & = \frac{1}{2\sqrt{\omega}} \biggl[\left(\mp \cos \varphi \partial_r  \pm \frac{1}{r} \sin \varphi
        \partial_{\varphi} + \omega r \cos \varphi \pm \cos \varphi \partial_r W \mp \frac{1}{r} \sin \varphi
        \partial_{\varphi} W\right) b_x \\
  & \quad + \left(\mp \sin \varphi \partial_r  \mp \frac{1}{r} \cos \varphi \partial_{\varphi} + \omega r 
        \sin \varphi \pm \sin \varphi \partial_r W \pm \frac{1}{r} \cos \varphi \partial_{\varphi} W\right) b_y 
        \biggr],
\end{split}  \label{eq:polar}
\end{equation}
\begin{equation}
\begin{split}
  K_0 & = K_{0,{\rm B}} + \Gamma, \qquad K_{\pm} = K_{\pm,{\rm B}} - \Gamma, \qquad Y = \frac{1}{2}
        \left(r \partial_r W + \bp_x b_x + \bp_y b_y - 1\right), \\
  K_{0,{\rm B}} & = D + \frac{1}{4} \omega r^2, \qquad K_{\pm,{\rm B}} = - D + \frac{1}{4} \omega r^2
        \mp \frac{1}{2} (r \partial_r + 1), \\
  D & = \frac{1}{4\omega} \left[- \partial_r^2 - \frac{1}{r} \partial_r - \frac{1}{r^2} \partial_{\varphi}^2 -
        \partial_r^2 W - \frac{1}{r} \partial_r W - \frac{1}{r^2} \partial_{\varphi}^2 W + (\partial_r W)^2
        + \frac{1}{r^2} (\partial_{\varphi} W)^2\right], \\
  \Gamma & = \frac{1}{2\omega} \biggl\{\biggl[\cos^2 \varphi \partial_r^2 W - \frac{2}{r} \sin \varphi
        \cos \varphi \partial_{r\varphi}^2 W + \frac{1}{r^2} \sin^2 \varphi \partial_{\varphi}^2 W + \frac{1}{r}
        \sin^2 \varphi \partial_r W \\
  & \quad + \frac{2}{r^2} \sin \varphi \cos \varphi \partial_{\varphi} W\biggr] \bp_x b_x + \biggl[\sin \varphi
        \cos \varphi \partial_r^2 W + \frac{1}{r} (\cos^2 \varphi - \sin^2 \varphi) \partial_{r\varphi}^2 W \\
  & \quad - \frac{1}{r^2} \sin \varphi \cos \varphi \partial_{\varphi}^2 W - \frac{1}{r} \sin \varphi \cos \varphi
        \partial_r W - \frac{1}{r^2} (\cos^2 \varphi - \sin^2 \varphi) \partial_{\varphi} W\biggr] \\
  & \quad \times (\bp_x b_y + \bp_y b_x) + \biggl[\sin^2 \varphi \partial_r^2 W + \frac{2}{r} \sin \varphi
        \cos \varphi \partial_{r\varphi}^2 W + \frac{1}{r^2} \cos^2 \varphi \partial_{\varphi}^2 W \\
  & \quad + \frac{1}{r} \cos^2 \varphi \partial_r W - \frac{2}{r^2} \sin \varphi \cos \varphi \partial_{\varphi} W
        \biggr] \bp_y b_y \biggr\}  
\end{split}  \label{eq:polar-bis}
\end{equation}
and 
\begin{equation}
  [r \partial_r, D] = - 2D, \qquad [r \partial_r, \Gamma] = - 2\Gamma, \qquad r \partial_r W = C.
  \label{eq:conditions-bis}
\end{equation}
\par
%
%--------------------------------------------------------------------------------------------------------------------
%
Conditions (\ref{eq:conditions-bis}) are readily satisfied if we assume that $W$ takes the form
\begin{equation*}
  W = C \ln r + F(\varphi),
\end{equation*}
where $F(\varphi)$ may be any (physically acceptable) function of $\varphi$.\par
%
%-------------------------------------------------------------------------------------------------------------------
%
In order to be relevant to the TTW Hamiltonians $H_k$, the realization (\ref{eq:polar}), (\ref{eq:polar-bis}) should be such that the bosonic part $4\omega K_{0,{\rm B}}$ of ${\cal H}^s$, defined in (\ref{eq:SUSY}), reduces to (\ref{eq:TTW}), which means that $D$ should be given by
\begin{equation*}
  D = \frac{1}{4\omega} \left\{- \partial_r^2 - \frac{1}{r} \partial_r - \frac{1}{r^2} \partial_{\varphi}^2 +
  \frac{k^2}{r^2} [a(a-1) \sec^2 k\varphi + b(b-1) \csc^2 k\varphi]\right\}. 
\end{equation*}
This will be so provided the function $F(\varphi)$ and the constant $C$ satisfy the Riccati equation
\begin{equation}
  - F'' + F^{\prime 2} + C^2 = k^2 [a(a-1) \sec^2 k\varphi + b(b-1) \csc^2 k\varphi].  \label{eq:riccati}
\end{equation}
A solution is easily found to be given by
\begin{equation}
  F(\varphi) = - a \ln \cos k\varphi - b \ln \sin k\varphi, \qquad C = - k(a+b).  \label{eq:F-C}
\end{equation}
It is worth observing here that to choose (\ref{eq:F-C}) among all the solutions of (\ref{eq:riccati}), we have been guided by the known results for Calogero-like Hamiltonians to be reviewed in section 5.\par
%
%------------------------------------------------------------------------------------------------------------------
%
On insering (\ref{eq:F-C}) in (\ref{eq:polar-bis}), we obtain that for the even generators the operators $\Gamma$ and $Y$ become
\begin{equation}
\begin{split}
  \Gamma & = \frac{k}{2\omega r^2}\biggl\{a \sec^2 k\varphi \biggl[\Bigl(\cos (k-2)\varphi \cos k\varphi
       + \frac{k}{2}(1 - \cos 2\varphi)\Bigr) \bp_x b_x \\  
  & \quad - \Bigl(\sin (k-2)\varphi \cos k\varphi + \frac{k}{2} \sin 2\varphi\Bigr) (\bp_x b_y + \bp_y b_x) + 
       \Bigl(- \cos (k-2)\varphi \cos k\varphi \\
  & \quad + \frac{k}{2}(1 + \cos 2\varphi)\Bigr) \bp_y b_y\biggr] + b \csc^2 k\varphi \biggl[\Bigl(\sin (k-2)
       \varphi \sin k\varphi + \frac{k}{2}(1 - \cos 2\varphi)\Bigr) \bp_x b_x \\ 
  & \quad + \Bigl(\cos (k-2)\varphi \sin k\varphi - \frac{k}{2} \sin 2\varphi\Bigr) (\bp_x b_y + \bp_y b_x) +
       \Bigl(- \sin (k-2)\varphi \sin k\varphi \\
  & \quad + \frac{k}{2}(1 + \cos 2\varphi)\Bigr) \bp_y b_y\biggr]\biggr\} 
\end{split}  \label{eq:Gamma}
\end{equation}
and
\begin{equation}
  Y = \frac{1}{2} [\bp_x b_x + \bp_y b_y - k(a+b) - 1],  \label{eq:Y}
\end{equation}
respectively. The supersymmetrized TTW Hamiltonian then assumes the form
\begin{equation}
  {\cal H}^s = H_{k,{\rm B}} + H_{k,{\rm F}}, \qquad H_{k,{\rm B}} = H_k, \qquad H_{k,{\rm F}} = 4\omega
  (\Gamma + Y).  \label{eq:super-TTW}
\end{equation}
\par
%
%-----------------------------------------------------------------------------------------------------------
%
{}Furthermore, the odd generators in (\ref{eq:polar}) turn out to be
\begin{equation}
\begin{split}
  V_{\pm} & = \frac{1}{2\sqrt{\omega}} \biggl[\biggl(\mp \cos \varphi \partial_r  \pm \frac{1}{r} \sin \varphi
        \partial_{\varphi} + \omega r \cos \varphi \pm \frac{ka}{r} \frac{\cos (k-1)\varphi}{\cos k\varphi} \\
  & \quad \pm \frac{kb}{r} \frac{\sin (k-1)\varphi}{\sin k\varphi}\biggr) \bp_x + \biggl(\mp \sin \varphi 
        \partial_r \mp \frac{1}{r} \cos \varphi \partial_{\varphi} + \omega r \sin \varphi \\
  & \quad \mp \frac{ka}{r} \frac{\sin (k-1)\varphi}{\cos k\varphi} \pm \frac{kb}{r} \frac{\cos (k-1)\varphi} 
        {\sin k\varphi}\biggr) \bp_y \biggr], \\
  W_{\pm} & = \frac{1}{2\sqrt{\omega}} \biggl[\biggl(\mp \cos \varphi \partial_r  \pm \frac{1}{r} \sin \varphi
        \partial_{\varphi} + \omega r \cos \varphi \mp \frac{ka}{r} \frac{\cos (k-1)\varphi}{\cos k\varphi} \\
  & \quad \mp \frac{kb}{r} \frac{\sin (k-1)\varphi}{\sin k\varphi}\biggr) b_x + \biggl(\mp \sin \varphi 
        \partial_r \mp \frac{1}{r} \cos \varphi \partial_{\varphi} + \omega r \sin \varphi \\
  & \quad \pm \frac{ka}{r} \frac{\sin (k-1)\varphi}{\cos k\varphi} \mp \frac{kb}{r} \frac{\cos (k-1)\varphi} 
        {\sin k\varphi}\biggr) b_y \biggr], 
\end{split}  \label{eq:V-W}
\end{equation}
with $W_+$ and $V_-$ providing the two supercharge operators for the supersymmetrized Hamiltonian (\ref{eq:super-TTW}) through equation (\ref{eq:SUSY}).\par
%
%-----------------------------------------------------------------------------------------------------------
%
By introducing two 'rotated' pairs of fermionic creation and annihilation operators $(\bbp_x, \bb_x)$ and $(\bbp_y, \bb_y)$, defined by
\begin{equation*}
  \bbp_x = \bp_x \cos \varphi + \bp_y \sin \varphi, \qquad \bbp_y = - \bp_x \sin \varphi + \bp_y \cos \varphi
\end{equation*}
and similarly for $\bb_x$, $\bb_y$, equations (\ref{eq:Gamma}), (\ref{eq:Y}) and (\ref{eq:V-W}) can be recast in a somewhat simpler form\footnote{It is worth observing here that these new fermionic operators should only be seen as a convenient tool to write the generators and their action on wavefunctions in a concise way since their dependence on $\varphi$ breaks the commutativity of bosonic and fermionic degrees of freedom.}
\begin{equation*}
\begin{split}
  \Gamma & = \frac{k}{2\omega r^2} \bigl\{a \bigl[\bbp_x \bb_x - \tan k\varphi \bigl(\bbp_x \bb_y 
       + \bbp_y \bb_x\bigr) + (k \sec^2 k\varphi - 1) \bbp_y \bb_y\bigr] \\
  & \quad + b \bigl[\bbp_x \bb_x + \cot k\varphi \bigl(\bbp_x \bb_y + \bbp_y \bb_x\bigr)
       + (k \csc^2 k\varphi - 1) \bbp_y \bb_y\bigr]\bigr\},
\end{split}
\end{equation*}
\begin{equation*}
  Y = \frac{1}{2} \bigl[\bbp_x \bb_x + \bbp_y \bb_y - k(a+b) - 1\bigr]
\end{equation*}
and
\begin{equation}
\begin{split}
  V_{\pm} & = \frac{1}{2\sqrt{\omega}} \biggl[\bbp_x \biggl(\mp \partial_r + \omega r \pm \frac{k(a+b)}{r}
         \biggr) \mp \bbp_y \frac{1}{r} (\partial_{\varphi} + ka \tan k\varphi - kb \cot k\varphi)\biggr], \\
  W_{\pm} & = \frac{1}{2\sqrt{\omega}} \biggl[\bb_x \biggl(\mp \partial_r + \omega r \mp \frac{k(a+b)}{r}
         \biggr) \mp \bb_y \frac{1}{r} (\partial_{\varphi} - ka \tan k\varphi + kb \cot k\varphi)\biggr].
\end{split}  \label{eq:V-W-bis}
\end{equation}
\par
%
%-------------------------------------------------------------------------------------------------------------------
%
In the next section, we will proceed to determine the action of the ${\rm osp}(2/2, \R)$ generators on the TTW Hamiltonian eigenstates after extending the latter with fermionic degrees of freedom.\par
%
%===============================================================
%
\section{\boldmath Irreducible representations of ${\rm osp}(2/2, \R)$ for the supersymmetrized TTW Hamiltonians}

In \cite{tremblay09}, it has been shown that $H_k$ is exactly solvable and satisfies the eigenvalue equation
\begin{equation*}
  H_k \Psi_{N,n}(r, \varphi) = E_{N,n} \Psi_{N,n}(r, \varphi), \qquad E_{N,n} = 2\omega [2N + (2n+a+b)k + 1],
\end{equation*}
where $N$, $n=0$, 1, 2,~\ldots. The wavefunctions can be written as
\begin{equation}
\begin{split}
  & \Psi_{N,n}(r, \varphi) = {\cal N}_{N,n} Z^{(2n+a+b)}_N (z) \Phi^{(a,b)}_n (\varphi), \\
  & Z^{(2n+a+b)}_N (z) = \left(\frac{z}{\omega}\right)^{\left(n + \frac{a+b}{2}\right) k} L^{((2n+a+b)k)}_N
          (z) e^{- \frac{1}{2} z}, \qquad z = \omega r^2, \\
  & \Phi^{(a,b)}_n (\varphi) = \cos^a k\varphi \sin^b k\varphi P^{\left(a - \frac{1}{2}, b - \frac{1}{2}\right)}_n
          (\xi), \qquad \xi = - \cos 2k\varphi,
\end{split}  \label{eq:w-f}
\end{equation}
in terms of Laguerre and Jacobi polynomials. In (\ref{eq:w-f}), ${\cal N}_{N,n}$ denotes a normalization constant, which can be easily calculated from some known properties of these polynomials \cite{gradshteyn} and is given by
\begin{equation*}
  {\cal N}_{N,n} = \bar{\cal N}_{N,n} {\cal N}^{(a,b)}_{n}
\end{equation*}
with
\begin{equation*}
\begin{split}
  & \bar{\cal N}_{N,n} = (-1)^N \left(\frac{2\, \omega^{(2n+a+b)k + 1} N!}{\Gamma(N + (2n+a+b)k +1)}\right)
        ^{1/2}, \\
  & {\cal N}^{(a,b)}_{n} = \left(\frac{2k\, n! (2n+a+b) \Gamma(a+b+n)}{\Gamma\left(a + n + \frac{1}{2}\right)
        \Gamma\left(b + n + \frac{1}{2}\right)}\right)^{1/2}.
\end{split}
\end{equation*}
Observe that the optional phase factor $(-1)^N$ in $\bar{\cal N}_{N,n}$ has been introduced to get positive matrix elements for the ${\rm sp}(2, \R)$ generators $K_{\pm}$ in conformity with the conventional choice.\par
%
%---------------------------------------------------------------------------------------------------------------
% 
After multiplication by the fermionic vacuum state $|0\rangle$ (i.e., $b_x |0\rangle = b_y |0\rangle = \bb_x |0\rangle = \bb_y |0\rangle = 0$), the wavefunctions (\ref{eq:w-f}) yield eigenstates of the supersymmetrized TTW Hamiltonian (\ref{eq:super-TTW}) with eigenvalues
\begin{equation*}
  {\cal E}_{N,n} = E_{N,n} - E_{0,0} = 4\omega (N + nk).
\end{equation*}
Such extended wavefunctions turn out to be also eigenstates of the ${\rm osp}(2/2, \R)$ weight generators $K_0$ and $Y$ corresponding to the eigenvalues $\tau + N$ and $q$, where
\begin{equation*}
  \tau = \left(n + \frac{a+b}{2}\right)k + \frac{1}{2}, \qquad q = - \frac{1}{2}[(a+b)k + 1].
\end{equation*}
\par
%
%----------------------------------------------------------------------------------------------------------------
%
All the states $\Psi_{N,n} |0\rangle$ with a definite value of $n$ (hence of $\tau$) and $N=0$, 1, 2,~\ldots\ belong to a ${\rm sp}(2, \R)$ lowest-weight state (LWS) irrep characterized by $\tau$ and will be denoted by
\begin{equation}
  |\tau, \tau + N, q\rangle = \Psi_{N,n} |0\rangle.  \label{eq:zero-fermion}
\end{equation}
They indeed satisfy the relations
\begin{equation*}
\begin{split}
  & K_0 |\tau, \tau + N, q\rangle = (\tau + N) |\tau, \tau + N, q\rangle, \\
  & K_+ |\tau, \tau + N, q\rangle = [(N + 1)(2\tau + N)]^{1/2} |\tau, \tau + N + 1, q\rangle, \\
  & K_- |\tau, \tau + N, q\rangle = [N(2\tau + N - 1)]^{1/2} |\tau, \tau + N - 1, q\rangle,
\end{split}
\end{equation*}
which can be easily checked by rewriting $K_{\pm}$ as
\begin{equation*}
  K_{\pm} = - \frac{1}{4\omega} H_k + \frac{1}{2} z \mp \left(z \partial_z + \frac{1}{2}\right) - \Gamma
\end{equation*}
and using some well-known properties of Laguerre polynomials \cite{gradshteyn}.\par
%
%-----------------------------------------------------------------------------------------------------------------------------
%
The odd generators $W_{\pm}$ annihilate the zero-fermion states (\ref{eq:zero-fermion}), whereas the remaining odd generators $V_{\pm}$ may lead to one- and two-fermion states. After some straightforward calculations, equations (\ref{eq:V-W-bis}) and (\ref{eq:w-f}) yield
\begin{equation}
\begin{split}
  & V_+ |\tau, \tau + N, q\rangle \\
  & \quad = \frac{{\cal N}_{N,n}}{\sqrt{z}} \Bigl\{\Bigl[ [N + (n+a+b)k + 1]
         Z^{(2n+a+b)}_N(z)  - (N+1) Z^{(2n+a+b)}_{N+1}(z)\Bigr] \\
  & \quad \times \Phi^{(a,b)}_n(\varphi) \bbp_x - (n+a+b)k Z^{(2n+a+b)}_N(z) 
         \Phi^{(a+1,b+1)}_{n-1}(\varphi) \bbp_y\Bigr\} |0\rangle, \\
  & V_- |\tau, \tau + N, q\rangle \\
  & \quad = \frac{{\cal N}_{N,n}}{\sqrt{z}} \Bigl\{\Bigl[ (N+nk) Z^{(2n+a+b)}_N(z)  
         - [N + (2n+a+b)k] Z^{(2n+a+b)}_{N-1}(z)\Bigr] \\
  & \quad \times \Phi^{(a,b)}_n(\varphi) \bbp_x + (n+a+b)k Z^{(2n+a+b)}_N(z) 
         \Phi^{(a+1,b+1)}_{n-1}(\varphi) \bbp_y\Bigr\} |0\rangle, \\
  & V_+ V_- |\tau, \tau + N, q\rangle = - V_- V_+ |\tau, \tau + N, q\rangle \\
  & \quad = {\cal N}_{N,n} (n+a+b)k Z^{(2n+a+b)}_N(z) \Phi^{(a+1,b+1)}_{n-1}(\varphi) \bbp_x \bbp_y
         |0\rangle.
\end{split}  \label{eq:V-appl}
\end{equation}
From (\ref{eq:def-1}), it is obvious that the three states in (\ref{eq:V-appl}) are eigenstates of $K_0$ and $Y$ with eigenvalues $\left(\tau + N + \frac{1}{2}, q + \frac{1}{2}\right)$, $\left(\tau + N - \frac{1}{2}, q + \frac{1}{2}\right)$ and $(\tau + N, q + 1)$, respectively.\par
%
%-------------------------------------------------------------------------------------------------------------------
% 
These one- and two-fermion states can be easily normalized by using the Hermiticity properties and anticommutation relations of the ${\rm osp}(2/2, \R)$ generators. The results read
\begin{equation}
\begin{split}
  & |+, \tau + N + \tfrac{1}{2}, q + \tfrac{1}{2}\rangle = [N + (n+a+b)k + 1]^{-1/2} V_+ |\tau, \tau + N, q
         \rangle, \\
  & |-, \tau + N - \tfrac{1}{2}, q + \tfrac{1}{2}\rangle = (N + nk)^{-1/2} V_- |\tau, \tau + N, q\rangle, \\
  & |\pm, \tau + N, q + 1\rangle = [n (n+a+b)k^2]^{-1/2} V_+ V_- |\tau, \tau + N, q\rangle.
\end{split}  \label{eq:normalized-fermionic}
\end{equation}
It can also be shown that the one-fermion states with the same eigenvalue of $K_0$ are not orthogonal and that their overlap is given by
\begin{equation*}
  \langle +, \tau + N - \tfrac{1}{2}, q + \tfrac{1}{2} | -, \tau + N - \tfrac{1}{2}, q + \tfrac{1}{2}\rangle =
  \left(\frac{N [N + (2n+a+b)k]}{[N + (n+a+b)k] (N+nk)}\right)^{1/2}
\end{equation*}
for $N=1$, 2~\ldots.\par
%
%-------------------------------------------------------------------------------------------------------------
%
The results obtained so far for one- and two-fermion states correspond to generic values of $N$ and $n$. To construct from them basis states for ${\rm sp}(2, \R)$ irreps we have to distinguish between vanishing and non-vanishing values of $n$.\par
%
%-------------------------------------------------------------------------------------------------------------------
%
{}For $n=0$, as a result of (\ref{eq:V-appl}) and of the properties $Z^{(a+b)}_{-1}(z) = \Phi^{(a+1,b+1)}_{-1}(\varphi) = 0$, it turns out that the ground state $|\tau, \tau, q\rangle = \Psi_{0,0} |0\rangle$ of ${\cal H}^s$ is annihilated not only by $K_-$ and $W_-$, but also by $V_-$. Hence it is an ${\rm osp}(2/2, \R)$ LWS and the states obtained from it by means of the raising generators form a basis  for a so-called atypical LWS irrep with $\tau = - q$ \cite{balantekin}. The latter is known to decompose into two ${\rm sp}(2, \R) \times {\rm so}(2)$ irreps characterized by $(\tau) (q)$ and $(\tau + \frac{1}{2}) (q + \frac{1}{2})$, respectively. This is confirmed by setting $n=0$ in the generic results (\ref{eq:V-appl}) and (\ref{eq:normalized-fermionic}). We indeed get
\begin{equation*}
  |-, \tau + N + \tfrac{1}{2}, q + \tfrac{1}{2}\rangle = |+, \tau + N + \tfrac{1}{2}, q + \tfrac{1}{2}\rangle, \qquad
  |\pm, \tau + N, q + 1\rangle = 0 
\end{equation*}
for $N=0$, 1, 2,~\ldots. In the $n=0$ case, we may therefore set
\begin{equation*}
  |\tau + \tfrac{1}{2}, \tau + N + \tfrac{1}{2}, q + \tfrac{1}{2}\rangle = |+, \tau + N + \tfrac{1}{2}, q + 
  \tfrac{1}{2}\rangle, \qquad N=0, 1, 2, \ldots.
\end{equation*}
It is worth observing that since $Q$ and $Q^{\dagger}$ in (\ref{eq:SUSY}) annihilate the ground state of ${\cal H}^s$, supersymmetry is unbroken.\par
%
%---------------------------------------------------------------------------------------------------------------
%
{}For non-vanishing values of $n$, the situation is more complicated since the three states in (\ref{eq:V-appl}) and (\ref{eq:normalized-fermionic}) are non-zero. It is however straightforward to show that the one-fermion states can be combined into basis states of two ${\rm sp}(2, \R)$ irreps, characterized by $\tau - \frac{1}{2}$ and $\tau + \frac{1}{2}$, respectively. The latter can be written as
\begin{equation*}
\begin{split}
  & |\tau - \tfrac{1}{2}, \tau + N - \tfrac{1}{2}, q + \tfrac{1}{2}\rangle = \alpha_N |-, \tau + N - \tfrac{1}{2}, 
        q + \tfrac{1}{2}\rangle + \beta_N |+, \tau + N - \tfrac{1}{2}, q + \tfrac{1}{2}\rangle, \\
  & |\tau + \tfrac{1}{2}, \tau + N + \tfrac{1}{2}, q + \tfrac{1}{2}\rangle = \gamma_N |-, \tau + N + \tfrac{1}{2}, 
        q + \tfrac{1}{2}\rangle + \delta_N |+, \tau + N + \tfrac{1}{2}, q + \tfrac{1}{2}\rangle,
\end{split}
\end{equation*}
with 
\begin{equation*}
\begin{split}
  & \alpha_N = \left(\frac{[N + (2n+a+b)k](N + nk)}{n (2n+a+b)k^2}\right)^{1/2}, \\ 
  & \beta_N = - \left(\frac{N[N + (n+a+b)k]}{n (2n+a+b)k^2}\right)^{1/2}, \\
  & \gamma_N = \left(\frac{(N + 1)(N + nk + 1)}{(n+a+b) (2n+a+b)k^2}\right)^{1/2}, \\ 
  & \delta_N = - \left(\frac{[N + (n+a+b)k + 1][N + (2n+a+b)k + 1]}{(n+a+b)(2n+a+b)k^2}\right)^{1/2}.        
\end{split}
\end{equation*}
On the other hand, the two-fermion states belong to a single ${\rm sp}(2, \R)$ LWS irrep specified by $\tau$:
\begin{equation*}
  |\tau, \tau + N, q + 1\rangle = |\pm, \tau + N, q + 1\rangle.
\end{equation*}
Hence, for any $n \ne 0$ there exists an ${\rm osp}(2/2, \R)$ irrep, which is not a LWS one and decomposes into four ${\rm sp}(2, \R) \times {\rm so}(2)$ irreps $(\tau) (q)$, $(\tau - \frac{1}{2}) (q + \frac{1}{2})$, $(\tau + \frac{1}{2}) (q + \frac{1}{2})$ and $(\tau) (q + 1)$.\par
%
%----------------------------------------------------------------------------------------------------------------------
%
We conclude that the eigenstates of any supersymmetrized TTW Hamiltonian may be separated into basis states of an infinite collection of ${\rm osp}(2/2, \R)$ irreps, each member of the set being characterized by a given value of the quantum number $n$ associated with the angular part of $H_k$ and determining the eigenvalue $(2n+a+b)^2 k^2$ of the first integral of motion $X_k$ \cite{tremblay09, cq10a}. The corresponding eigenvalues of the second- and third-order Casimir operators \cite{frappat}
\begin{equation*}
\begin{split}
  C_2 &= K_0(K_0-1) - Y(Y+1) - K_+ K_- + V_- W_+ - V_+ W_-, \\
  C_3 &= (K_0 + Y)(K_0 - Y -1)(Y + \tfrac{1}{2}) - (Y + \tfrac{1}{2}) K_+ K_- + \tfrac{1}{2}[K_- V_+ - (K_0 
         - 3Y) V_-] W_+ \\
  & \quad + \tfrac{1}{2}[K_+ V_- - (K_0 + 3Y) V_+] W_-
\end{split}
\end{equation*}
are given by $n(n+a+b)k^2$ and $- \frac{1}{2} (a+b)n(n+a+b)k^3$, respectively. They vanish for $n=0$ in agreement with a known property of atypical LWS irreps. It is worth mentioning that similar relationships between Casimir operators and the angular part of Hamiltonians have already been observed in other contexts \cite{ghosh04, hakobyan}.\par
%
%=====================================================================
%
\section{\boldmath The $k=1$, 2 and 3 cases}

The purpose of this section is to establish some connections between the outcomes of section 3 and some known results corresponding to $k=1$, 2 and 3. For simplicity's sake, we will restrict ourselves here to the explicit relations obtained for ${\cal H}^s$ and $Q$, defined in (\ref{eq:SUSY}), as the remaining ${\rm osp}(2/2, \R)$ generators can be easily dealt with in the same way.\par
%
%+++++++++++++++++++++++++++++++++++++++++++++++++++++++++++++
%
\subsection{\boldmath SW model ($k=1$)}

The SW Hamiltonian \cite{fris, winternitz}
\begin{equation*}
  H_1 = - \partial_x^2 - \partial_y^2 + \omega^2 (x^2 + y^2) + \frac{a(a-1)}{x^2} + \frac{b(b-1)}{y^2}
\end{equation*}
is usually written in cartesian coordinates, wherein it is also separable.\par
%
%-------------------------------------------------------------------------------------------------------
%
On setting $k=1$ in (\ref{eq:Gamma}), (\ref{eq:Y}), (\ref{eq:super-TTW}) and (\ref{eq:V-W}) and going back from $r$, $\varphi$ to $x$, $y$, we directly get
\begin{equation*}
\begin{split}
  {\cal H}^s & = - \partial_x^2 - \partial_y^2 + \omega^2 (x^2 + y^2) + \frac{a^2}{x^2} + \frac{b^2}{y^2}
        + 2\omega (\bp_x b_x + \bp_y b_y) + \frac{a}{x^2} [\bp_x, b_x] \\
  & \quad + \frac{b}{y^2} [\bp_y, b_y] - 2\omega (a+b+1), \\
  Q & = \left(- \partial_x + \omega x - \frac{a}{x}\right) b_x + \left(- \partial_y + \omega y - \frac{b}{y}\right)
        b_y. 
\end{split}
\end{equation*}
Similar results would be obtained either by supersymmetrizing separately the two one-dimensional cartesian Hamiltonians or by setting $W = - a \ln |x| - b \ln |y|$ in equations (\ref{eq:Calogero-1}) and (\ref{eq:Calogero-2}).\par
%
%+++++++++++++++++++++++++++++++++++++++++++++++++++++++++++++++++++
%
\subsection{\boldmath $BC_2$ model ($k=2$)}

The Hamiltonian of the $BC_2$ model is given by \cite{olsha83}
\begin{equation*}
  H_2 = - \partial_x^2 - \partial_y^2 + \omega^2 (x^2 + y^2) + 2a(a-1) \left(\frac{1}{(x-y)^2} + 
  \frac{1}{(x+y)^2}\right) + b(b-1) \left(\frac{1}{x^2} + \frac{1}{y^2}\right).   
\end{equation*}
\par
%
%------------------------------------------------------------------------------------------------------------------
%
{}For $k=2$, equations (\ref{eq:Gamma}), (\ref{eq:Y}), (\ref{eq:super-TTW}) and (\ref{eq:V-W}) easily yield
\begin{equation*}
\begin{split}
  {\cal H}^s & = - \partial_x^2 - \partial_y^2 + \omega^2 (x^2 + y^2) + 2a^2 \left(\frac{1}{(x-y)^2} + 
       \frac{1}{(x+y)^2}\right) + b^2 \left(\frac{1}{x^2} + \frac{1}{y^2}\right) \\
  & \quad + 2\omega (\bp_x b_x + \bp_y b_y)  + a \left(\frac{1}{(x-y)^2} [\bp_x - \bp_y, b_x - b_y] + 
       \frac{1}{(x+y)^2} [\bp_x + \bp_y, b_x + b_y]\right) \\
  & \quad + b \left(\frac{1}{x^2} [\bp_x, b_x] + \frac{1}{y^2} [\bp_y, b_y]\right) - 2\omega (2a+2b+1), \\
  Q & = \left[- \partial_x + \omega x - a \left(\frac{1}{x-y} + \frac{1}{x+y}\right) - \frac{b}{x}\right] b_x \\
  & \quad + \left[- \partial_y + \omega y + a \left(\frac{1}{x-y} - \frac{1}{x+y}\right) - \frac{b}{y}\right] b_y,  
\end{split}
\end{equation*}
which would also result from section 2 by using $W = - a (\ln |x-y] + \ln |x+y|) - b (\ln |x| + \ln |y|)$, in agreement with \cite{brink98}.\par
%
%+++++++++++++++++++++++++++++++++++++++++++++++++++++++++++++++
%
\subsection{\boldmath CMW model ($k=3$)}

In contrast with the two previous cases, the comparison of the results obtained in section 3 for $k=3$ with those of the supersymmetrized CMW model is more involved because the latter is a three-particle system, which can only be interpreted as a planar problem after eliminating the centre-of-mass motion. On starting from \cite{wolfes}
\begin{equation*}
  H_{\rm CMW} = \sum_i (- \partial_i^2 + \omega^2 x_i^2) + a(a-1) \sum_{\substack{i,j \\ i\ne j}} 
  \frac{1}{x_{ij}^2} + 3b(b-1) \sum_{\substack{i,j \\ i\ne j}} \frac{1}{y_{ij}^2},
\end{equation*}
where $x_{ij} = x_i - x_j$ ($i \ne j$), $y_{ij} = x_i + x_j - 2x_k$ ($i \ne j \ne k \ne i$) and all indices run over 1, 2, 3, we indeed get
\begin{equation*}
  H_{\rm CMW} = H_{\rm rel} + H_{\rm cm}, \qquad H_{\rm rel} = H_3, \qquad H_{\rm cm} = -\partial_X^2
  + \omega^2 X^2,
\end{equation*}
by setting $r \cos \varphi = x_{12}/\sqrt{2}$, $r \sin \varphi = y_{12}/\sqrt{6}$ and $X = (x_1 + x_2 + x_3)/\sqrt{3}$.\par
%
%-----------------------------------------------------------------------------------------------------------------------
%
The supersymmetrized CMW Hamiltonian and the corresponding supercharge are obtained by inserting
\begin{equation*}
  W = - \frac{a}{2} \sum_{\substack{i,j \\ i\ne j}} \ln |x_{ij}| - \frac{b}{2} \sum_{\substack{i,j \\ i\ne j}} 
  \ln |y_{ij}|  
\end{equation*}
in (\ref{eq:Calogero-1}) and (\ref{eq:Calogero-2}) and they are given by
\begin{equation}
\begin{split}
  {\cal H}^s_{\rm CMW} & = \sum_i (- \partial_i^2 + \omega^2 x_i^2) + a^2 \sum_{\substack{i,j \\ i\ne j}} 
       \frac{1}{x_{ij}^2} + 3b^2 \sum_{\substack{i,j \\ i\ne j}} \frac{1}{y_{ij}^2} + 2\omega \sum_i \bp_i b_i \\
  & \quad + a \sum_{\substack{i,j \\ i\ne j}} \frac{1}{x_{ij}^2} [\bp_i, b_i - b_j] + b 
       \sum_{\substack{i,j,k \\ i\ne j\ne k\ne i}} \frac{1}{y_{ij}^2} ([\bp_i, b_i + b_j - 2b_k] \\
  & \quad - [\bp_k, b_i + b_j - 2b_k]) - 3\omega (2a+2b+1), \\
  Q_{\rm CMW} & = \sum_i \left[- \partial_i + \omega x_i - a \sum_{\substack{j \\ j\ne i}} \frac{1}{x_{ij}} 
       - b \left(\sum_{\substack{j \\ j\ne i}} \frac{1}{y_{ij}} - \sum_{\substack{j,k \\ i\ne j\ne k\ne i}} 
       \frac{1}{y_{jk}}\right)\right] b_i,
\end{split}  \label{eq:super-CMW}
\end{equation}
in terms of three pairs of fermionic creation and annihilation operators $\bp_i$, $b_i$, $i=1$, 2, 3.\par
%
%--------------------------------------------------------------------------------------------------------------------
%
Let us now make for the latter the same kind of orthogonal transformation as that performed for the coordinates,
\begin{equation*}
  \bp_x = \frac{1}{\sqrt{2}} (\bp_1 - \bp_2), \qquad \bp_y =  \frac{1}{\sqrt{6}} (\bp_1 + \bp_2 - 2 \bp_3), 
  \qquad \bp_X =  \frac{1}{\sqrt{3}} (\bp_1 + \bp_2 + \bp_3)
\end{equation*}
and similarly for the annihilation operators. Then, after some calculations, equation (\ref{eq:super-CMW}) can be rewritten as
\begin{equation*}
  {\cal H}^s_{\rm CMW} = {\cal H}^s_{\rm rel} + {\cal H}^s_{\rm cm}, \qquad Q_{\rm CMW} = Q_{\rm rel} 
  + Q_{\rm cm},
\end{equation*}
where
\begin{equation}
\begin{split}
  {\cal H}^s_{\rm rel} & = - \partial_r^2 - \frac{1}{r} \partial_r - \frac{1}{r^2} \partial_{\varphi}^2 + \omega^2
        r^2 + \frac{1}{r^2 \cos^2 \varphi} a (a + [\bp_x, b_x]) \\
  & \quad + \frac{1}{r^2 \cos^2 \left(\varphi - \frac{2\pi}{3}\right)} a \left(a + \frac{1}{4} \left[\sqrt{3} \bp_y
        - \bp_x, \sqrt{3} b_y - b_x\right]\right) \\
  & \quad + \frac{1}{r^2 \cos^2 \left(\varphi - \frac{4\pi}{3}\right)} a \left(a + \frac{1}{4} \left[\sqrt{3} \bp_y
        + \bp_x, \sqrt{3} b_y + b_x\right]\right) \\
  & \quad + \frac{1}{r^2 \sin^2 \varphi} b (b + [\bp_y, b_y]) \\
  & \quad + \frac{1}{r^2 \sin^2 \left(\varphi - \frac{2\pi}{3}\right)} b \left(b + \frac{1}{4} \left[\sqrt{3} \bp_x
        + \bp_y, \sqrt{3} b_x + b_y\right]\right) \\
  & \quad + \frac{1}{r^2 \sin^2 \left(\varphi - \frac{4\pi}{3}\right)} b \left(b + \frac{1}{4} \left[\sqrt{3} \bp_x
        - \bp_y, \sqrt{3} b_x - b_y\right]\right) \\
  & \quad + 2\omega (\bp_x b_x + \bp_y b_y) - 2\omega (3a+3b+1),  
\end{split}  \label{eq:super-rel}
\end{equation}
\begin{equation}
\begin{split}
  Q_{\rm rel} & = \left(- \cos \varphi \partial_r + \frac{1}{r} \sin \varphi \partial_{\varphi} + \omega r 
        \cos \varphi - \frac{3a}{r} \frac{\cos 2\varphi}{\cos 3\varphi} - \frac{3b}{r} 
        \frac{\sin 2\varphi}{\sin 3\varphi}\right) b_x \\ 
  & \quad + \left(- \sin \varphi \partial_r - \frac{1}{r} \cos \varphi \partial_{\varphi} + \omega r 
        \sin \varphi + \frac{3a}{r} \frac{\sin 2\varphi}{\cos 3\varphi} - \frac{3b}{r} 
        \frac{\cos 2\varphi}{\sin 3\varphi}\right) b_y
\end{split}  \label{eq:Q-rel}
\end{equation}
and
\begin{equation*}
\begin{split}
  & {\cal H}^s_{\rm cm} = - \partial_X^2 + \omega^2 X^2 + 2\omega (\bp_X b_X - \tfrac{1}{2}), \\
  & Q_{\rm cm} = (- \partial_X + \omega X) b_X.
\end{split}
\end{equation*}
\par
%
%-----------------------------------------------------------------------------------------------------------------
%
It is immediately clear that $Q_{\rm rel}$ in (\ref{eq:Q-rel}) coincides with $2 \sqrt{\omega}\, W_+$ obtained by setting $k=3$ in (\ref{eq:V-W}). To show that ${\cal H}^s_{\rm rel}$ in (\ref{eq:super-rel}) also reduces to the supersymmetrized TTW Hamiltonian with $k=3$, given in (\ref{eq:Gamma}), (\ref{eq:Y}) and (\ref{eq:super-TTW}), requires some work, but this can be easily done by employing well-known trigonometric identities similar to those used in \cite{cq10a, cq10b}, thereby completing the comparison.\par
%
%====================================================================
%
\section{Conclusion}

In the present paper, we have obtained, for any positive real $k$, a ${\cal N} = 2$ supersymmetric extension
${\cal H}^s$ of the TTW Hamiltonians $H_k$ on a plane, which generalizes the known ones for $k=1$, 2 and 3. Such a supersymmetrized TTW Hamiltonian has an ${\rm osp}(2/2, \R)$ dynamical superalgebra, whose irreps have been shown to be characterized by the quantum number $n$ specifying either the angular wavefunctions of $H_k$ or the eigenvalues of its first integral of motion $X_k$. Bases for these irreps have been explicitly built and the irrep containing the ground state of ${\cal H}^s$ has been identified as an atypical LWS one.\par
%
%-------------------------------------------------------------------------------------------------------------
%
Several interesting questions are raised by the results obtained in this work, such as the feasibility of constructing higher $\cal N$ extensions  of $H_k$, the possible existence of hidden nonlinear  supersymmetries and the relation between the present ${\cal N}=2$ supersymmetric extension and the previously considered one based on the dihedral group $D_{2k}$. We hope to come back to some of these issues in forthcoming publications.\par
%
%=============================================================
% 

\end{document}